\documentclass[aps,prc,twocolumn,superscriptaddress,showpacs]{revtex4-1}   
\usepackage{lineno}

\usepackage{amsmath}
\usepackage{amsfonts}
\usepackage{multirow}
\usepackage{verbatim}
\usepackage{color}     
\usepackage{array}
\usepackage{dcolumn}
\usepackage{xspace}
\usepackage{textcomp}
\usepackage{graphicx}
\usepackage{gensymb}
\usepackage[colorlinks,linkcolor=blue,citecolor=blue,urlcolor=blue]{hyperref}
\graphicspath{{figs/}}

\newcommand{\geant}{\textsc{geant4}\xspace}
\newcommand{\Duke}{%
	Department of Physics, Duke University, Durham, North Carolina 27708-0308, USA}
\newcommand{\TUNL}{%
	Triangle Universities Nuclear Laboratory, Durham, North Carolina 27708-0308, USA}
\newcommand{\NCCU}{%
	Department of Mathematics and Physics, North Carolina Central University, Durham, North Carolina, 27707, USA}
\newcommand{\GWUINT}{%
	Institute for Nuclear Studies, Department of Physics, The George Washington University, Washington DC 20052, USA}
\newcommand{\UNC}{%
	University of North Carolina at Chapel Hill, Chapel Hill, North Carolina 27516, USA}
\newcommand{\UKY}{%
	Department of Physics and Astronomy, University of Kentucky, Lexington, Kentucky 40506, USA}
\newcommand{\USask}{%
	Department of Physics and Engineering Physics, University of Saskatchewan, Saskatoon, Saskatchewan, S7N 5E2, Canada}
\newcommand{\NGC}{%
	Department of Physics, University of North Georgia, Dahlonega, Georgia 30597, USA}
\newcommand{\JMU}{%
	Department of Physics and Astronomy, James Madison University, Harrisonburg, Virginia 22807, USA}
\newcommand{\NCState}{%
    Department of Physics, North Carolina State University, Raleigh, North Carolina 27695, USA}

\begin{document}
 
\title{Compton scattering from \texorpdfstring{$^4$He}{} at the TUNL \texorpdfstring{HI$\gamma$S}{} facility}

\author{X.~Li}
\email[]{xiaqing.li@duke.edu}
\affiliation{\Duke{}}
\affiliation{\TUNL{}}

\author{M.W.~Ahmed}
\affiliation{\Duke{}}
\affiliation{\TUNL{}}
\affiliation{\NCCU{}}

\author{A.~Banu}
\affiliation{\JMU{}}

\author{C.~Bartram}
\affiliation{\TUNL{}}
\affiliation{\UNC{}}

\author{B.~Crowe}
\affiliation{\TUNL{}}
\affiliation{\NCCU{}}

\author{E.J.~Downie}
\affiliation{\GWUINT{}}

\author{M.~Emamian}
\affiliation{\TUNL{}}

\author{G.~Feldman}
\affiliation{\GWUINT{}}

\author{H.~Gao}
\affiliation{\Duke{}}
\affiliation{\TUNL{}}

\author{D.~Godagama}
\affiliation{\UKY{}}

\author{H.W.~Grie{\ss}hammer}
\affiliation{\GWUINT{}}
\affiliation{\Duke{}}

\author{C.R.~Howell}
\affiliation{\Duke{}}
\affiliation{\TUNL{}}

\author{H.J.~Karwowski}
\affiliation{\TUNL{}}
\affiliation{\UNC{}}

\author{D.P.~Kendellen}
\affiliation{\Duke{}}
\affiliation{\TUNL{}}

\author{M.A.~Kovash}
\affiliation{\UKY{}}

\author{K.K.H.~Leung}
\affiliation{\TUNL{}}
\affiliation{\NCState{}}

\author{D.~Markoff}
\affiliation{\TUNL{}}
\affiliation{\NCCU{}}

\author{S.~Mikhailov}
\affiliation{\TUNL{}}

\author{R.E.~Pywell}
\affiliation{\USask{}}

\author{M.H.~Sikora}
\affiliation{\GWUINT{}}
\affiliation{\TUNL{}}

\author{J.A.~Silano}
\affiliation{\TUNL{}}
\affiliation{\UNC{}}

\author{R.S.~Sosa}
\affiliation{\NCCU{}}

\author{M.C.~Spraker}
\affiliation{\NGC{}}

\author{G.~Swift}
\affiliation{\TUNL{}}

\author{P.~Wallace}
\affiliation{\TUNL{}}

\author{H.R.~Weller}
\affiliation{\Duke{}}
\affiliation{\TUNL{}}

\author{C.S.~Whisnant}
\affiliation{\JMU{}}

\author{Y.K.~Wu}
\affiliation{\Duke{}}
\affiliation{\TUNL{}}

\author{Z.W.~Zhao}
\affiliation{\Duke{}}
\affiliation{\TUNL{}}

\begin{abstract}
Differential cross sections for elastic Compton scattering from $^4$He have been measured with high statistical precision at the High Intensity $\gamma$-ray Source at laboratory scattering angles of $55\degree$, $90\degree$, and $125\degree$ using a quasi-monoenergetic photon beam with a weighted mean energy value of 81.3\,MeV. The results are compared to previous measurements and similar fore-aft asymmetry in the angular distribution of the differential cross sections is observed. This experimental work is expected to strongly motivate the development of effective-field-theory calculations of Compton scattering from $^4$He to fully interpret the data.
\end{abstract}

\maketitle

\section{Introduction}
\label{sec:intro}
Nucleon polarizabilities are of fundamental importance for understanding the dynamics of the internal structure of nucleons. The static electric and magnetic dipole polarizabilities, $\alpha_{E1}$ and $\beta_{M1}$, characterize the response of the nucleon to external electromagnetic stimulus by relating the strength of the induced electric and magnetic dipole moments of the nucleon to the applied field. Decades-long endeavors have been devoted to studying the static polarizabilities of nucleons experimentally and theoretically~\cite{Schumacher05,Griesshammer12,Hagelstein15}. Nuclear Compton scattering is a powerful tool to access the nucleon polarizabilities, where incident real photons apply an electromagnetic field to the nucleon and induce multipole radiation by displacing the charges and currents inside the nucleon. In the past decade, effective field theories (EFTs) have proven to be successful theoretical frameworks to describe such processes as well as to predict and extract static nucleon polarizabilities from the low-energy Compton scattering data~\cite{Griesshammer12}. Values of $\alpha_{E1}$ and $\beta_{M1}$ of the proton have been successfully extracted from Compton scattering experiments using liquid hydrogen targets. With $\alpha_{E1}^p+\beta_{M1}^p$ constrained by the Baldin sum rule (BSR), the latest EFT fit to the global database of proton Compton scattering gives~\cite{McGovern12,Griesshammer15}
\begin{equation}
\begin{split}
\alpha_{E1}^p &= 10.65\pm0.35_{\rm stat}\pm0.2_{\rm BSR}\pm0.3_{\rm theo},\\
\beta_{M1}^p &= 3.15\mp0.35_{\rm stat}\pm0.2_{\rm BSR}\mp0.3_{\rm theo},
\end{split}
\end{equation}
where the polarizabilities are given here and throughout this paper in units of $10^{-4}\,\rm fm^3$. The neutron polarizabilities, in contrast, are less well determined due to the lack of free neutron targets and the small Thomson cross sections of the neutron due to the fact that the neutron is uncharged~\cite{Schumacher05,Myers12,Myers:2014ace}.

Light nuclear targets, such as liquid deuterium~\cite{Hornidge99,Lundin02,Myers15,Myers:2014ace}, liquid $^4$He~\cite{Sikora17,Fuhrberg95,Proff99}, and $^6$Li~\cite{Myers12,Myers14}, can be utilized as effective neutron targets to extract neutron polarizabilities. After accounting for the binding effects, these isoscalar targets allow for the extraction of the isoscalar-averaged polarizabilities of the proton and neutron. By subtracting the better-known proton results, the neutron polarizabilities can be obtained. Indeed, the group at the MAX~IV Laboratory in Lund reported the most recent EFT extraction of the neutron polarizabilities from the world data of elastic deuteron Compton scattering as~\cite{Myers:2014ace}
\begin{equation}
\begin{split}
\alpha_{E1}^n &= 11.55\pm1.25_{\rm stat}\pm0.2_{\rm BSR}\pm0.8_{\rm theo}, \\
\beta_{M1}^n &= 3.65\mp1.25_{\rm stat}\pm0.2_{\rm BSR}\mp0.8_{\rm theo},
\end{split}
\end{equation}
with the BSR constraint applied. 

Although no model-independent calculation currently exists for nuclei with mass number $A>3$, light nuclei of higher masses are still advantageous in real Compton scattering experiments. Their cross sections are much larger both due to a higher atomic number $Z$ compared to the deuteron, and due to the fact that the meson-exchange currents play a larger role. In particular, $^4$He is a favorable candidate among isoscalar targets because, unlike $^6$Li, its description is well within the reach of modern high-accuracy theoretical approaches. Our data will show that the cross section of Compton scattering from $^4$He is approximately a factor of 6 to 8 larger than that from the deuteron. This substantial enhancement in the cross section enables high statistics measurements. Besides, this confirms the idea that Compton scattering cross sections do not only scale with $Z$ but are also sensitive to the amount of nuclear binding. The Compton scattering cross section, in turn, is related to the number of charged meson-exchange pairs in the target nucleus, see Ref.~\cite{Margaryan18} for details. Additionally, because the first inelastic channel is the $^4$He($\gamma,p$)$^3$H reaction with a $Q$ value of 19.8\,MeV and there are no bound excited states below that energy, elastic Compton scattering from $^4$He can be more easily distinguished from inelastic scattering compared to the deuteron with 2.2\,MeV binding energy. Now that the theory for $^3$He has been explored extensively~\cite{Margaryan18,Shukla:2018rzp,Shukla:2008zc}, one can expect a full theoretical calculation of Compton scattering from $^4$He as a next step. The first precise measurement of the $^4$He Compton scattering cross section was successfully performed at the High Intensity $\gamma$-ray Source (HI$\gamma$S) facility at an incident photon energy of 61\,MeV with high statistical accuracy and well-controlled systematic uncertainties~\cite{Sikora17}. A measurement at a higher photon energy is then motivated in order to obtain greater sensitivity to nucleon polarizabilities and to stimulate the development of EFT calculations of $^4$He Compton scattering to fully interpret the data.

In this paper, we report a new high-precision measurement of the cross section of elastic Compton scattering from $^4$He at a weighted mean incident photon beam energy of 81.3\,MeV. The results are compared to the previous $^4$He Compton scattering data and are discussed in the context of their significance to the extraction of the nucleon polarizabilities.

\section{Experimental Setup}
\label{sec:experiment}
The experiment was performed at HI$\gamma$S at the Triangle Universities Nuclear Laboratory (TUNL)~\cite{Weller09}. The HI$\gamma$S facility utilizes a storage-ring based free-electron laser (FEL) to produce intense, quasi-monoenergetic, and nearly 100\% circularly and linearly polarized $\gamma$-ray beams via Compton backscattering~\cite{Yan:2019bru,Wu:2015hta,Wu:2006zzc}. The $\gamma$-ray beam pulses have a width of about 300\,ps FWHM and are separated by 179\,ns. These features of the HI$\gamma$S photon beam lead to detector energy spectra which are much cleaner and simpler to interpret compared to Compton scattering experiments that use tagged bremsstrahlung beams.

The $\gamma$-ray beam was collimated by a lead collimator with a circular opening of 25.4\,mm diameter located 52.8\,m downstream from the electron-photon collision point. The $\gamma$-ray beam energy was determined from the set-point energy of the storage ring and the measured wavelength of the FEL beam. In this experiment, the calculated energy spectrum of the $\gamma$-ray beam incident on the liquid $^4$He target was peaked at around 85\,MeV with an estimated rms uncertainty of about 1\%~\cite{WYing}. In our experiment an accurate determination of the energy distribution of the $\gamma$-ray beam at the lower energy end was not possible due to the insertion of a set of apertured copper absorbers inside the FEL cavity, which preferably attenuated low energy $\gamma$ rays at larger angles. Instead a study of the beam energy profile was performed to obtain the weighted mean energy of the incident $\gamma$-ray beams. The weighted mean energy was the average energy weighted by the numbers of photons with different energies in the beam. Details of the determination of the weighted mean beam energy are discussed in Sec.~\ref{sec:analysis}. The $\gamma$-ray beam flux was continuously monitored by a system composed of five thin plastic scintillators with an aluminum radiator inserted after the second scintillator~\cite{Pywell09}. This system was located about 70\,cm downstream from the end of the collimator and about 12\,m upstream of the target. The charged particles produced in the radiator were detected to determine the photon beam intensity. The efficiency of the beam flux monitor was measured at low photon fluxes using a large NaI(Tl) detector located downstream of the target and was corrected for multiple hits at high photon rates~\cite{Pywell09}. For the present experiment, the on-target intensity of the circularly polarized photon beams was $\approx10^7\,\gamma/s$.

The beam was scattered from a cryogenic liquid $^4$He target~\cite{Kendellen16}. The near-cylindrical target cell was 20\,cm in length and approximately 4\,cm in diameter. The walls and end windows of the cell were made from 0.125-mm-thick Kapton foil. The cylindrical axis of the target cell was aligned along the beam axis. The target cell was located inside an aluminum vacuum can and two aluminum radiation shields. Two 0.125-mm-thick Kapton windows were installed on the vacuum can, and gaps were cut in the aluminum shields to allow the photon beam to enter and exit the cryogenic target. The liquid $^4$He target was maintained at 3.4\,K with a target thickness of $(4.17\pm0.04) \times 10^{23}$\,nuclei/cm$^2$. During production runs, the cell was periodically emptied for background measurements.

The Compton scattered photons were detected by an array of eight NaI(Tl) detectors placed at scattering angles of $\theta = 55\degree$, 90$\degree$, and 125$\degree$ in the laboratory frame. Each detector consisted of a cylindrical NaI(Tl) core of diameter 25.4\,cm surrounded by eight annular segments of 7.5-cm-thick NaI(Tl) crystals. The lengths of the core NaI crystals ranged from 25.4 to 30.5\,cm. The annular segments were used as an anticoincidence shield to veto the cosmic-ray background. A 15-cm-thick lead collimator was installed at the front face of each detector to define the acceptance cone. The conical aperture of the lead shield was filled with boron-doped paraffin wax to reduce neutron background. 

\begin{figure}[!htb]
\begin{center}
\includegraphics[width=1\columnwidth]{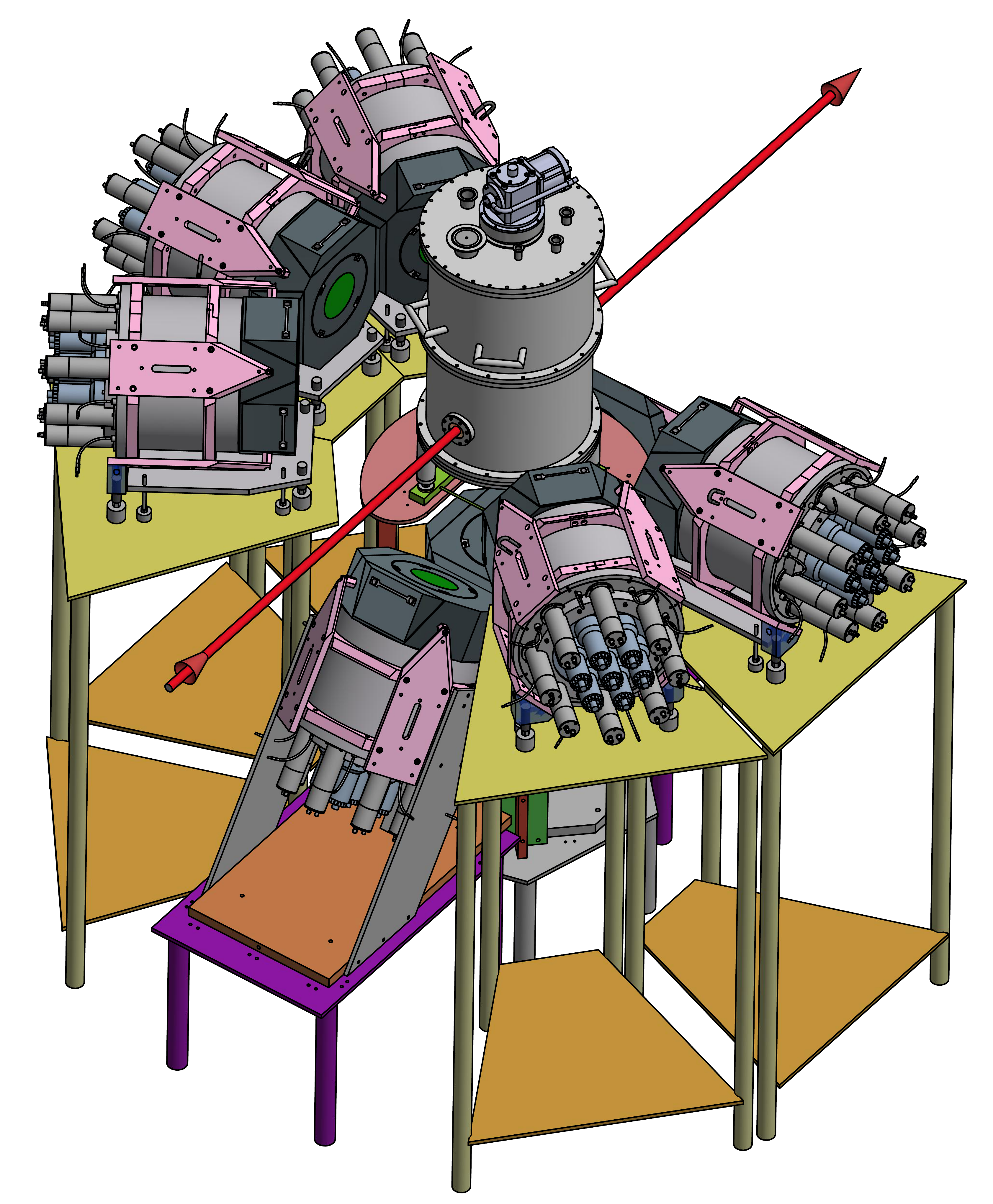}
\caption{(Color online) Schematic of the experimental apparatus showing the layout of the cryogenic target and the array of NaI(Tl) detectors. The photon beam is incident from the lower left side of the figure. The target cell is contained inside the aluminum vacuum can. \label{fig:Setup}}
\end{center}
\end{figure}

The layout of the detectors and the cryogenic target is shown in Fig.~\ref{fig:Setup}. For this experiment, five detectors (one at $\theta = 55\degree$, two at $\theta = 90\degree$, the other two at $\theta = 125\degree$) were placed on the tables with the axes of the detectors and the target aligned in the same horizontal plane. The other three detectors were placed beneath the beam axis pointing towards the target, located at $\theta = 55\degree$, $90\degree$, and $125\degree$, respectively. The geometry of the experimental apparatus was surveyed to a precision of 0.5\,mm and the results were incorporated into a \geant~\cite{Agostinelli02} simulation to determine the effective solid angle of each detector. The effective solid angle accounts for the geometric effects due to the extended target and the finite acceptance of the detectors, as well as the attenuation of scattered photons in the target cell and the cryostat. With the front faces of the detector collimator apertures placed about 58\,cm from the target center, the simulated effective solid angles ranged from 63.4 to 66.9\,msr.

\section{Data Analysis}
\label{sec:analysis}
\begin{figure}[!htp]
\includegraphics[width=1\columnwidth]{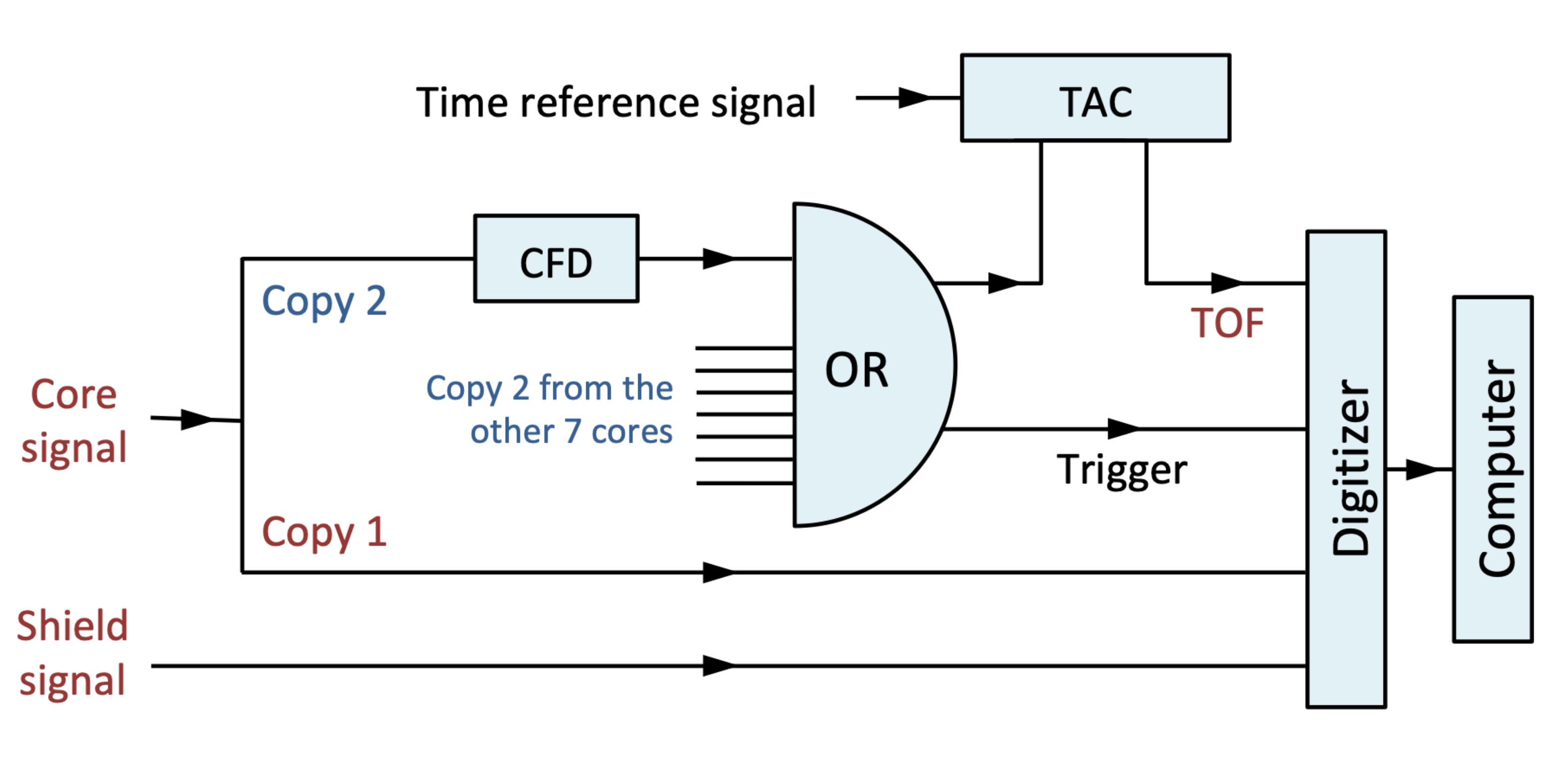}
\caption{(Color online) Simplified diagram of the data acquisition system. \label{fig:Daq}}
\end{figure}

Figure~\ref{fig:Daq} illustrates the simplified flow chart of the data acquisition system for this experiment. One copy of the core signal was recorded by the digitizer, while a second copy was used to generate the trigger for the data acquisition system. After passing a hardware threshold of about 10\,MeV, which was set using a constant fraction discriminator (CFD), a logical OR of all core signals was formed to trigger the digitizer. For each event trigger, in addition to recording the waveform of the signal from the core detector that generated the trigger, the waveform of the combined signal from the eight NaI shield segments associated with this core NaI detector and the time of flight (TOF) of the detected $\gamma$-ray events produced by the signal from a time-to-amplitude converter (TAC) were digitized. The TOF was defined as the time difference between an event trigger and the next reference signal of the electron beam pulse from the accelerator every 179\,ns. The energy deposition in the core and shield detectors was extracted from the integral of the pulse shape, while the TOF was obtained from the peak-sensed amplitude of the waveform from the TAC. 

\begin{figure}[!htp]
\includegraphics[angle=0,width=1\columnwidth]{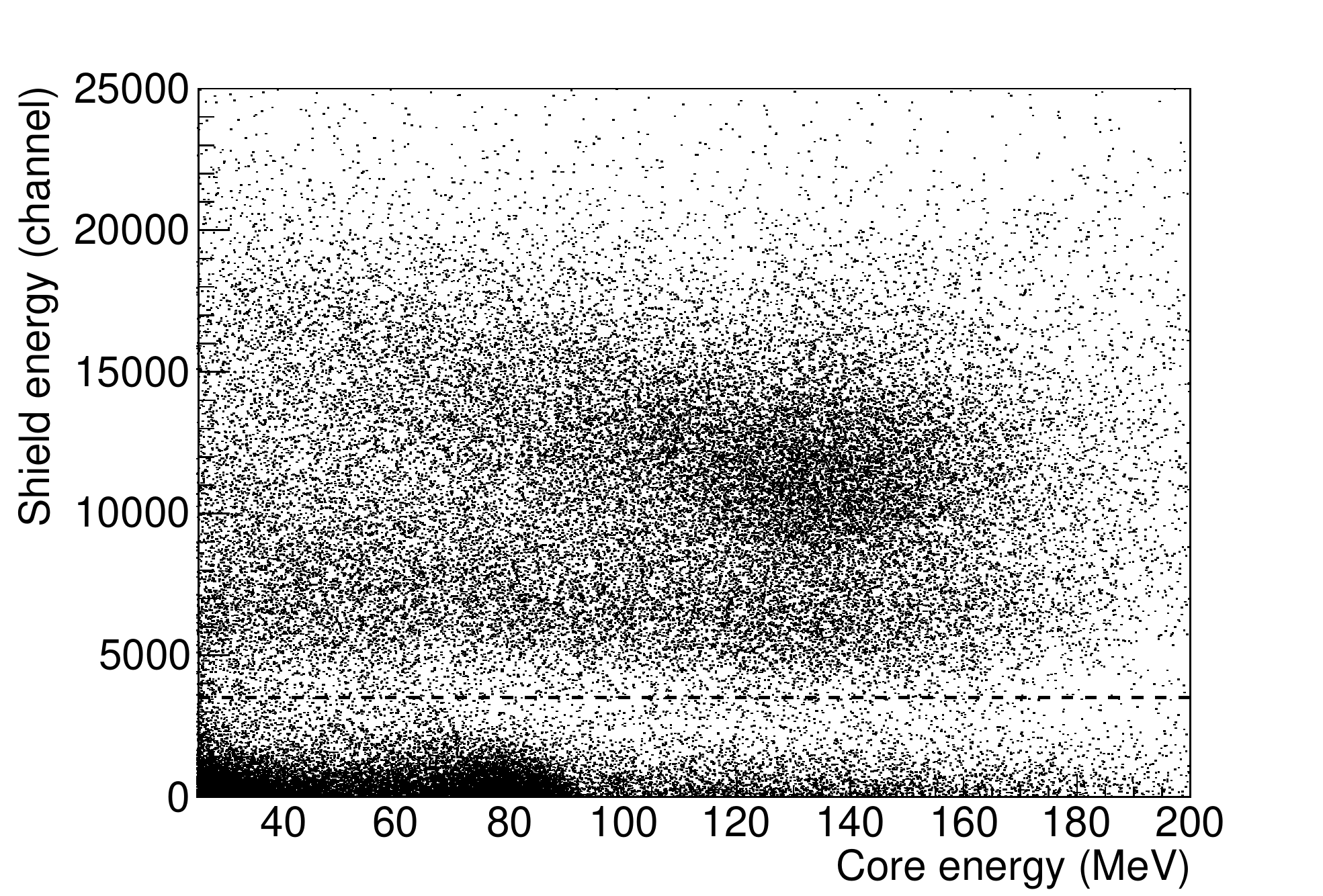}
\caption{2D spectrum showing energy deposition in the shield detector versus energy deposition in the core detector. An apparent gap around the dashed line is observed between the cosmic-ray events (above the dashed line) and the Compton-scattering events (below the dashed line). The shield energy cut is placed at the dashed line. 
\label{fig:shield2d}}
\end{figure}

\begin{figure}[!htp]
\includegraphics[angle=0,width=1\columnwidth]{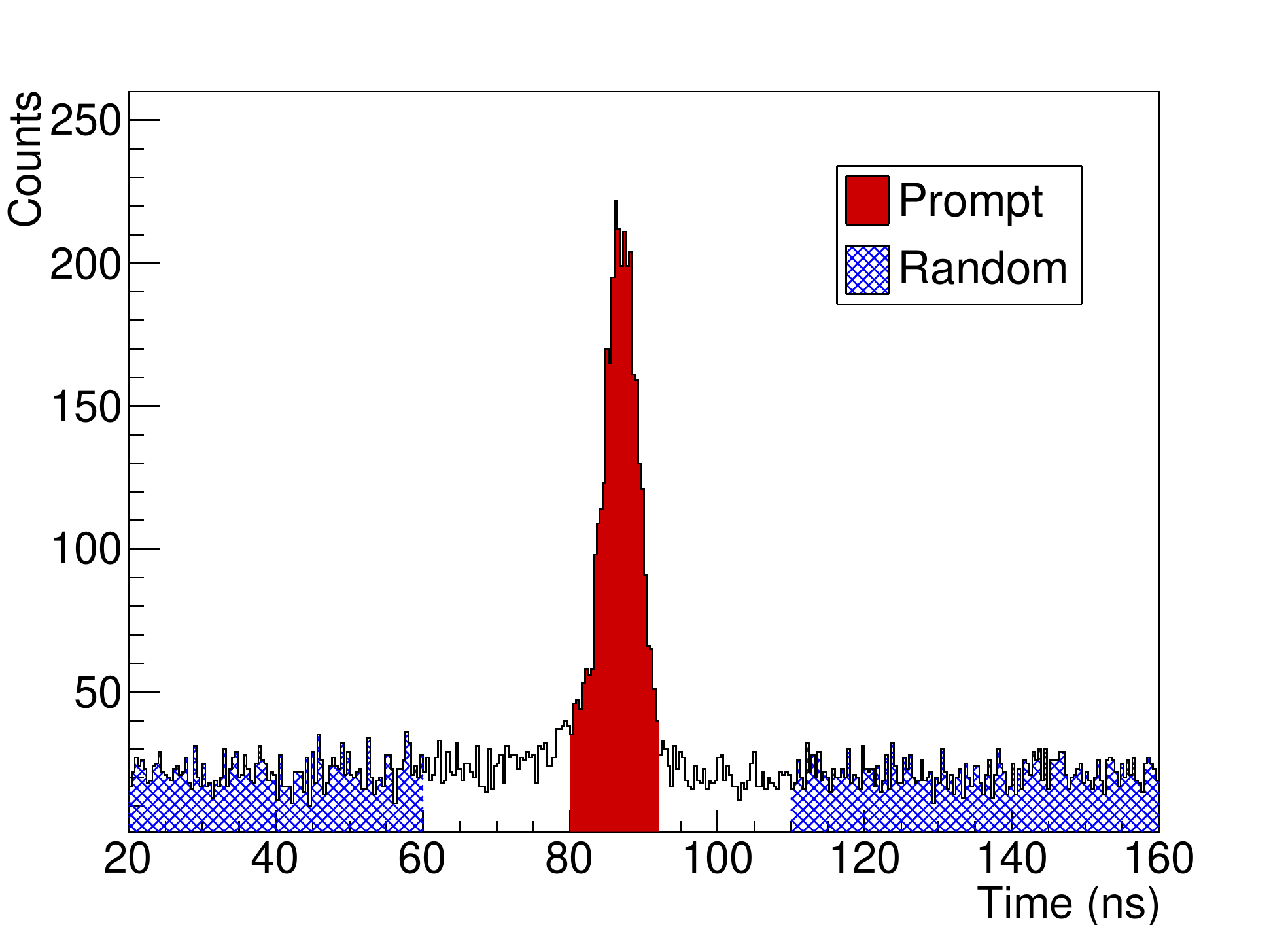}
\caption{(Color online) Time-of-flight spectrum showing prompt and random regions with shield-energy cut applied. The bin width is 0.4\,ns. 
\label{fig:ToF}}
\end{figure}

\begin{figure}[!htp]
\includegraphics[width=1\columnwidth]{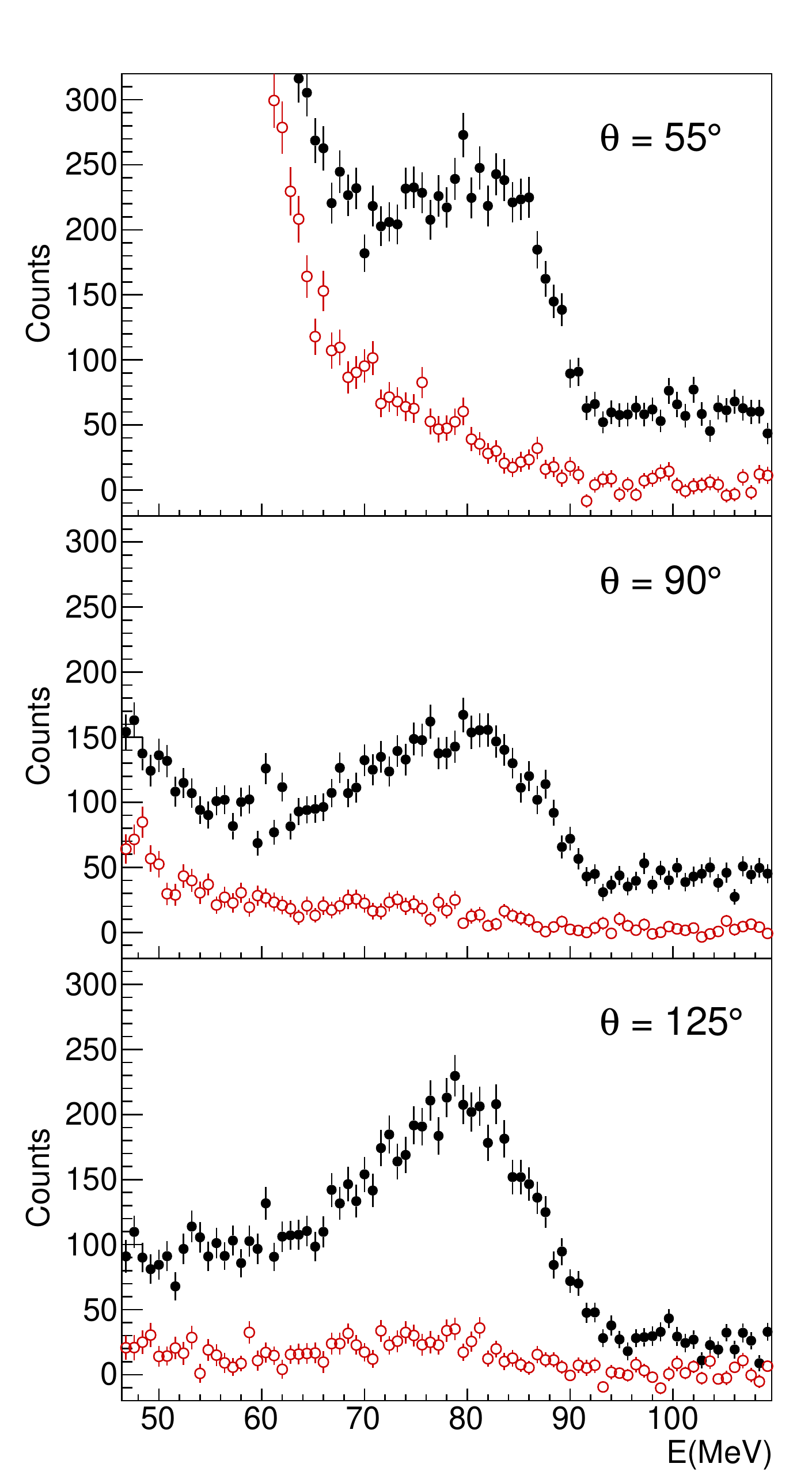}
\caption{(Color online) Representative energy spectra for full (closed dot) and empty (open dot) targets at $\theta$ = 55$\degree$, 90$\degree$, and 125$\degree$. The bin width is 0.8\,MeV. The empty-target spectra have been normalized to the number of incident photons for the full-target spectra. 
\label{fig:FullEmpty}}
\end{figure}

Cosmic-ray events were the major source of background in this experiment and could be rejected by employing two methods. First, the energy spectra of the anticoincidence shield detectors were analyzed to suppress such background. Due to the lead collimator in front of each detector, events scattered from the target deposited energy in the shields primarily through electromagnetic shower loss from the core crystal. In contrast, high-energy muons produced by cosmic rays traversed the detector and were minimum ionizing. This significant difference in the shield-energy spectra of the Compton scattered photons and cosmic muons enabled a cut on shield energy (Fig.~\ref{fig:shield2d}) to veto the cosmic-ray background without affecting the Compton scattering events. Secondly, the time structure of the $\gamma$-ray beam produced a clear prompt timing peak (Fig.~\ref{fig:ToF}) for the beam-produced events, allowing for a timing cut to select beam-related scattering events. The shield-energy cut and timing cut together removed over 99\% of the cosmic-ray events within the region of interest (ROI) in the energy spectrum. The background from time-uncorrelated (random) events appeared as a uniform distribution in the TOF spectrum in Fig.~\ref{fig:ToF}. After applying both cuts, the remaining background from residual random events was removed by sampling the energy spectrum from the random region and subtracting it from the energy spectrum in the prompt region after normalizing to the relative widths of the timing windows. The above analysis was performed on both full- and empty-target data. Typical energy spectra from the analysis at the three scattering angles are shown in Fig.~\ref{fig:FullEmpty}. For each detector, the empty-target energy spectrum was subtracted from the full-target energy spectrum after scaling to the number of incident photons to obtain the final energy spectrum.

\begin{figure}[!htp]
\includegraphics[width=1\columnwidth]{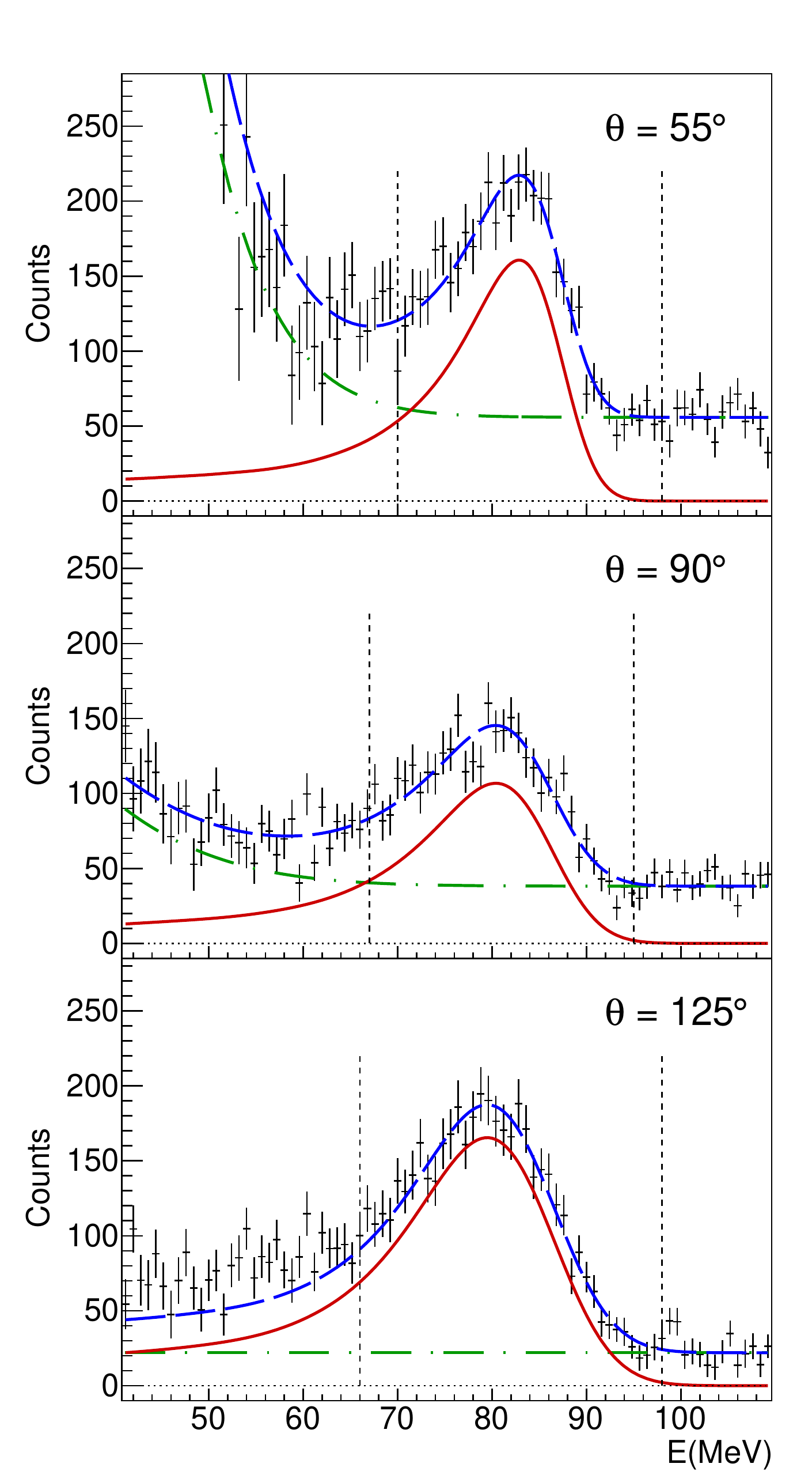}
\caption{(Color online) Representative final energy spectra at $\theta$ = 55$\degree$, 90$\degree$, and 125$\degree$ with empty-target events removed. The bin width is 0.8\,MeV. The fit to the data (dashed curve) consists of the electromagnetic background (dot-dashed curve) and the {\geant} simulated detector response function (solid curve). At $\theta$ = 55$\degree$ and 90$\degree$, the backgrounds are the sum of an exponential low-energy contribution accounting for the atomic scattering and a constant background resulting from the electron-beam-induced bremsstrahlung photons. At $\theta$ = 125$\degree$, the background is free from the exponential low-energy component and therefore includes the constant contribution only. The vertical dashed lines indicate the ROI used to obtain the yield at each angle. 
\label{fig:FitSpectra}}
\end{figure}

\begin{figure}[!htb]
\begin{center}
\includegraphics[angle=0,width=1\columnwidth]{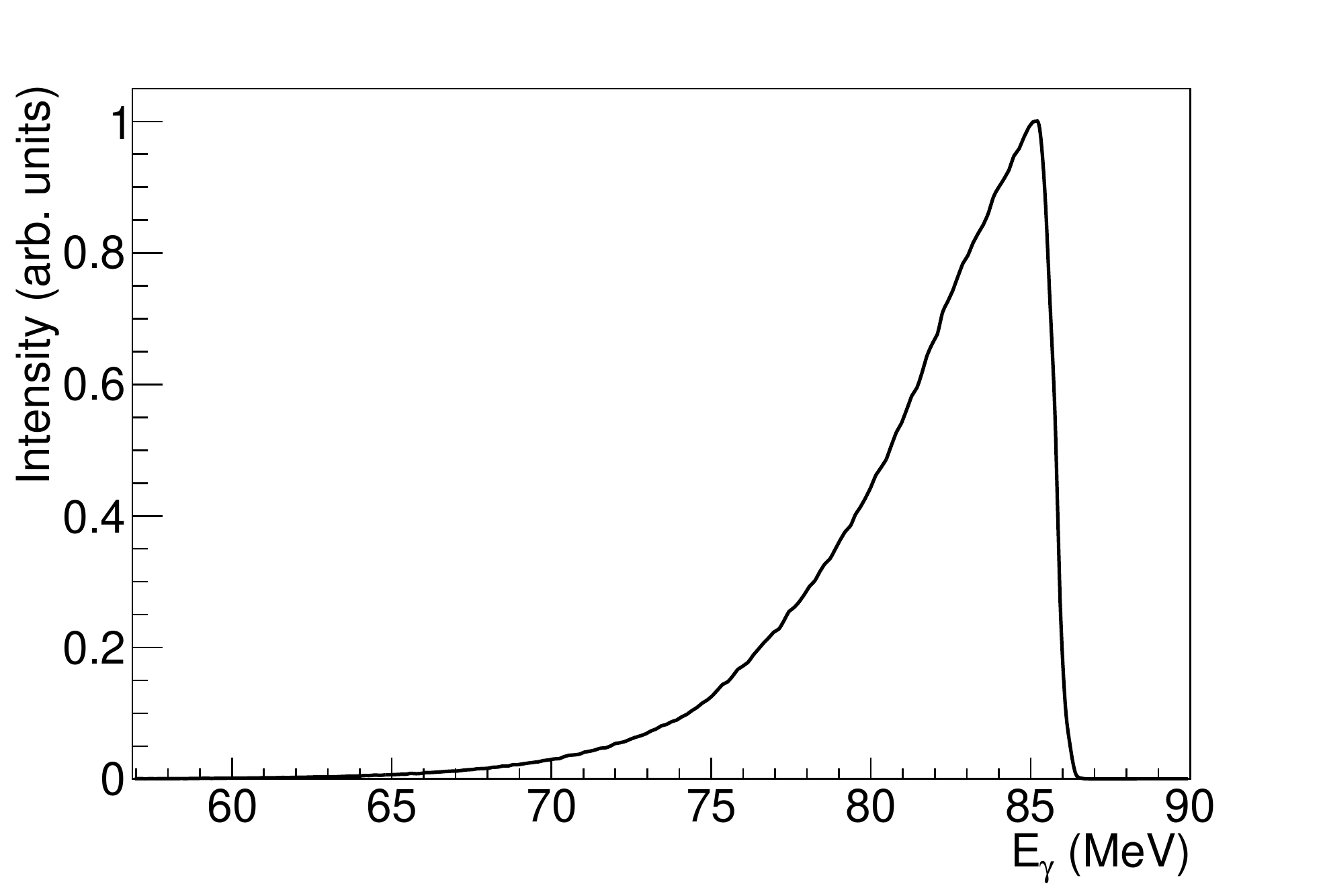}
\caption{The reconstructed effective beam energy profile on the target for a $\gamma$-ray beam produced by an electron beam energy of 975\,MeV in the storage ring and a laser beam of about 192 nm in the FEL optical cavity. This energy distribution is peaked at 85.1\,MeV with a FWHM of 5.5\,MeV (6.5\%). The weighted mean value is 81.3\,MeV.
\label{fig:Beam}}
\end{center}
\end{figure}

\begin{figure}[!htp]
\includegraphics[angle=0,width=1\columnwidth]{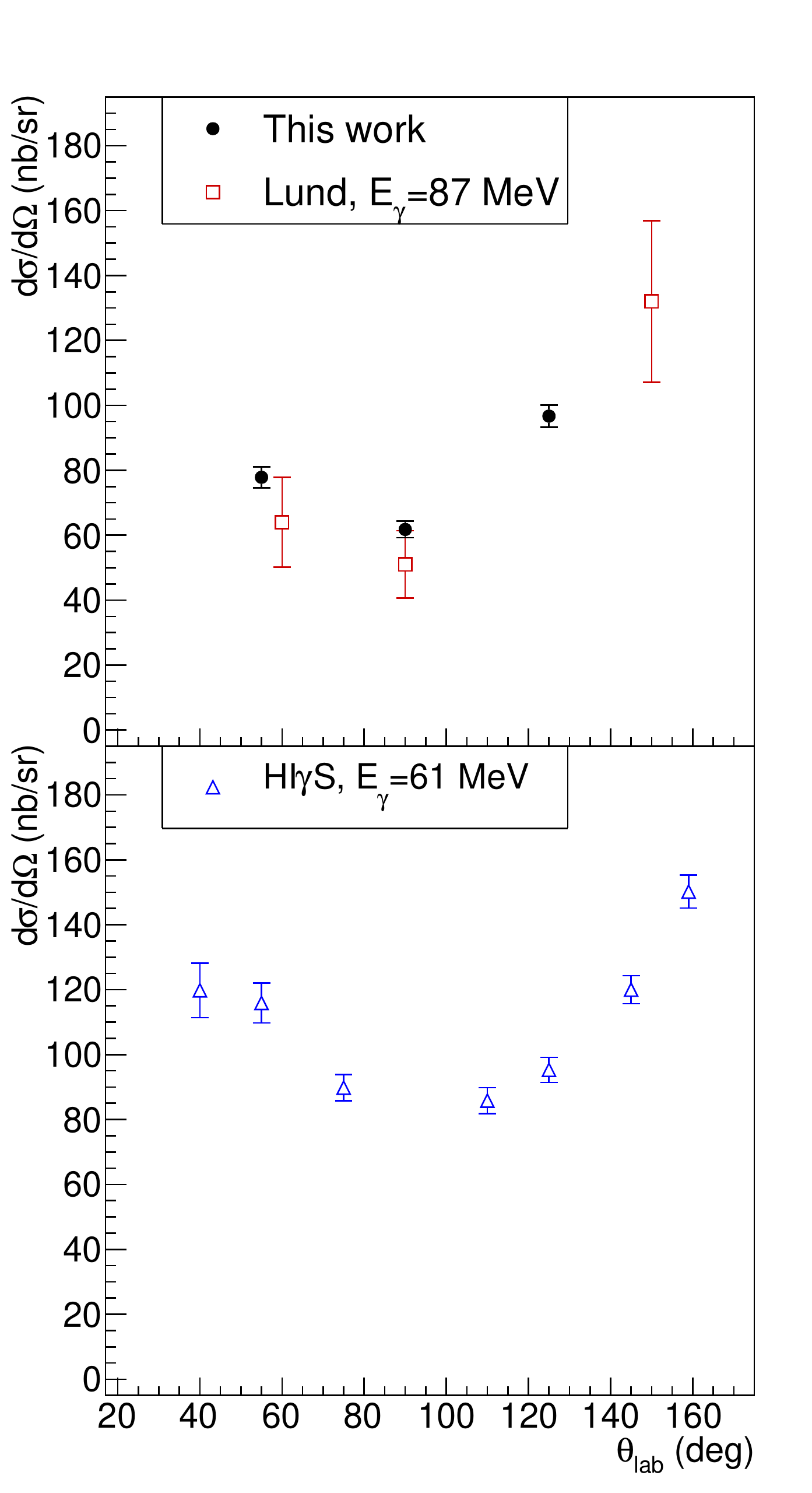}
\caption{(Color online) Compton scattering cross sections of $^4$He reported in the current work (circles) compared to the results from Lund (squares, $E_\gamma=$ 87\,MeV)~\cite{Fuhrberg95} and the previous measurement at HI$\gamma$S (triangles, $E_\gamma=$ 61\,MeV)~\cite{Sikora17}. The error bars shown are the statistical and systematic uncertainties added in quadrature. 
\label{fig:CrossSections}}
\end{figure}

Typical final energy spectra at the three scattering angles are shown in Fig.~\ref{fig:FitSpectra}. The final energy spectrum of each detector was fitted with the detector response function obtained from the aforementioned {\geant} simulation of the full experimental apparatus. Photons were generated from the target cell and the energy deposited in the detectors was recorded. The energy of the simulated scattered photon $E_\gamma^{\prime}$ was
\begin{equation}
     E_\gamma^{\prime} = \frac{E_\gamma}{1 + \frac{E_\gamma}{M}(1-\cos\theta)},
\end{equation}
where $E_\gamma$ is the incident photon energy sampled from the effective beam energy profile, $\theta$ is the laboratory scattering angle, and $M$ is the mass of the $^4$He nucleus. The simulated detector response function was fitted to the final energy spectrum with a Gaussian smearing convolution accounting for the intrinsic detector resolution. Due to the lack of a direct method to determine the beam energy profile in the entire distribution range, the effective beam energy profile was determined using the following strategy. A series of samples of beam energy profiles were calculated with a set of beam radii on the target, corresponding to various energy spreads. Each sample was incorporated into the aforementioned simulation to obtain the corresponding detector response function. As the detector response function was fitted to data, the resulting fitting parameter, particularly the width of the Gaussian smearing function representing the intrinsic detector resolution, was evaluated. Only those samples resulting in physically reasonable values of the fitting parameter were taken into account to estimate the weighted mean beam energy, which gave a range of 80.2\,MeV to 84.0\,MeV among the samples. The effective beam energy profile used in the cross-section extraction is shown in Fig.~\ref{fig:Beam} with a FWHM of 5.5\,MeV (6.5\%) and a weighted mean energy value of 81.3\,MeV. A flat background resulting from the scattering of bremsstrahlung photons produced from the 1\,GeV electrons in the storage ring was fitted simultaneously to the final energy spectrum. For the forward-angle detectors, the background from atomic Compton scattering was prominent in the low-energy region and was fitted with an exponential function in addition to the flat background. Typical line-shape fitting for the final energy spectra at forward and backward angles is shown in Fig.~\ref{fig:FitSpectra}. The fitted backgrounds were subtracted from the final energy spectrum, and the number of events was counted within the ROI in the elastic peak as indicated by the vertical dashed lines in Fig.~\ref{fig:FitSpectra}. The yield was extracted by scaling the summed number with an efficiency factor defined as the fraction of the fitted response function within the ROI. The aforementioned shield-energy and timing cuts were chosen such that the maximum yield was obtained while the uncertainty of the yield was minimized. As the lowest-energy breakup reaction $^4$He($\gamma,p$)$^3$H is 19.8\,MeV away from the elastic channel, the inelastic contribution is not expected to contaminate the ROI. The yield was corrected for the absorption of the incident photon beam in the target and normalized to the number of incident photons, target thickness, and effective solid angles to calculate the differential cross sections. 

The systematic uncertainties in this experiment were grouped into two categories and are listed in Table~\ref{tab:Systematic}. The first was the point-to-point systematic uncertainty that varied from each datum. This type of uncertainty reflected the effects of placing cuts in the energy and timing spectra of each detector in yield extraction. Values of the point-to-point uncertainties were determined by slightly varying the boundaries of ROI, the windows for the line-shape fitting, the timing cut, and the shield-energy cut. The contribution from the last item was negligible compared to the others. The second type was an overall normalization uncertainty that applied equally to all data. This overall normalization uncertainty included the contributions from the number of incident photons and the target thickness. The uncertainty from using different effective beam profiles from the selected samples was found negligible. Also, the uncertainty from the effective solid angles was evaluated and found negligible due to the high precision in the geometry survey of the experimental apparatus and the small statistical uncertainty of the simulation of the detector system. The contributions of the systematic uncertainties were summed in quadrature to obtain the total systematic uncertainty.

\begin{table}[]
\begin{center}
\caption{Systematic uncertainties.}
\begin{tabular*}{1\columnwidth}{@{\extracolsep{\fill}} l l r }
\hline\hline\noalign{\smallskip}
Type            & Source                        & Value \\
\noalign{\smallskip}\hline\noalign{\smallskip}
Point-to-point  & Timing cut            & 0.3\%--1.7\% \\
                & Line-shape fit window & 0.4\%--1.9\% \\
                & ROI                   & 0.3\%--1.3\% \\
\noalign{\smallskip}\noalign{\smallskip}
Normalization   & Number of incident photons    & 2.0\% \\
                & Target thickness              & 1.0\% \\
\noalign{\smallskip}\noalign{\smallskip}
Total           &                               & 2.5\%--3.5\%  \\
\noalign{\smallskip}\hline\hline
\end{tabular*}
\label{tab:Systematic}
\end{center}
\end{table}

\begin{table}[!htp]
\begin{center}
\caption{The Compton scattering cross section of $^4$He at a weighted mean beam energy of 81.3\,MeV measured by the eight NaI(Tl) detectors used in our experiment. The last three columns list the statistical, point-to-point systematic, and total systematic uncertainties.}
\begin{tabular*}{1\columnwidth}{@{\extracolsep{\fill}} l c c c c c}
\hline\hline\noalign{\smallskip}
$\theta_{\text{Lab}}$  &$\phi$ & $d\sigma/d\Omega$ & Stat & Point-to-point &Total Syst \\
\noalign{\smallskip}
 &  & (nb/sr) & (nb/sr) &Syst (nb/sr) &(nb/sr)\\
\noalign{\smallskip}\hline\noalign{\smallskip}
 $55\degree$    & $0\degree$    & 75.1  &$\pm2.6$   &$\pm1.1$  &$\pm2.0$ \\
 $55\degree$    & $270\degree$  & 81.6  &$\pm2.8$   &$\pm2.2$  &$\pm2.9$ \\
\noalign{\smallskip}\noalign{\smallskip}
 $90\degree$    & $0\degree$    & 58.8  &$\pm2.2$   &$\pm1.0$  &$\pm1.6$ \\
 $90\degree$    & $180\degree$  & 66.7  &$\pm2.7$   &$\pm1.0$  &$\pm1.8$ \\
 $90\degree$    & $270\degree$  & 61.8  &$\pm2.2$   &$\pm0.7$  &$\pm1.6$ \\
\noalign{\smallskip}\noalign{\smallskip}
 $125\degree$   & $0\degree$    & 90.6  &$\pm2.6$   &$\pm1.2$  &$\pm2.4$ \\
 $125\degree$   & $180\degree$  & 97.2  &$\pm2.6$   &$\pm1.2$  &$\pm2.5$ \\
 $125\degree$   & $270\degree$  & 102.1  &$\pm2.4$   &$\pm1.1$  &$\pm2.5$ \\
\noalign{\smallskip}\hline\hline
\end{tabular*}
\label{tab:CrossSections}
\end{center}
\end{table}

\begin{table}[!htp]
\begin{center}
\caption{The averaged Compton scattering cross section of $^4$He measured at weighted mean beam energy 81.3\,MeV at the three angles of our experimental setup. These data are plotted in Fig.~\ref{fig:CrossSections}. The third column lists the statistical uncertainties. The last two columns list the point-to-point and total systematic uncertainties.}
\begin{tabular*}{1\columnwidth}{@{\extracolsep{\fill}} l c c c c}
\hline\hline\noalign{\smallskip}
$\theta_{\text{Lab}}$  & $d\sigma/d\Omega$ & Stat & Point-to-point & Total Syst \\
\noalign{\smallskip}
 & (nb/sr) & (nb/sr) &Syst (nb/sr) &(nb/sr)\\
\noalign{\smallskip}\hline\noalign{\smallskip}
 $55\degree$ &   78.0 &$\pm1.9$ &$\pm1.7$ &$\pm2.4$\\
 $90\degree$ &   61.9 &$\pm1.3$ &$\pm0.9$ &$\pm1.6$\\
 $125\degree$ &  97.0 &$\pm1.5$ &$\pm1.2$ &$\pm2.5$\\
\noalign{\smallskip}\hline\hline
\end{tabular*}
\label{tab:Results}
\end{center}
\end{table}

\section{Results and Discussion}
\label{sec:results}
The differential cross section extracted for each detector is listed in Table~\ref{tab:CrossSections}. For different detectors at the same angle $\theta$, the cross sections are overall in good agreement with each other, although spreads among different azimuthal angles are more pronounced at $125\degree$. Part of the spreads can be accounted for once systematic uncertainties are taken into account. However, systematic aspects due to detection variations among detectors are difficult to take into account in our analysis but can likely contribute to the observed spreads. The cross section value at each scattering angle in the end is assigned as the weighted average by the statistical uncertainties. The reduced $\chi^{2}$ is calculated at each scattering angle with respect to the weighted average value. The calculated reduced $\chi^{2}$ values are $2.07$, $2.17$, and $4.36$ at $\theta = 55\degree$, $90\degree$, and $125\degree$, respectively. The final results are plotted in Fig.~\ref{fig:CrossSections} and listed in Table~\ref{tab:Results}. The elastic Compton scattering data from Lund~\cite{Fuhrberg95} at an incident photon energy of 87\,MeV and the previous HI$\gamma$S measurement~\cite{Sikora17} at 61\,MeV are also shown in Fig.~\ref{fig:CrossSections}. The results of this work are in good agreement with the Lund results. Particularly, the present data follow the same fore-aft asymmetry in the angular distribution as the Lund data, with a strong backward peaking that is evident in both data sets. This asymmetry is distinct from the 61-MeV HI$\gamma$S data and indicates a strong sensitivity to sub-nuclear effects, including the nucleon polarizabilities. An accurate theoretical calculation of the reaction, currently lacking, is needed to explain this behavior as well as to extract the values for the isoscalar polarizabilities from these data. Such a calculation has already been done for Compton scattering from $^3$He using an EFT framework~\cite{Margaryan18,Shukla:2018rzp,Shukla:2008zc}, therefore the prospects for such a treatment for $^4$He are very promising. The high statistical accuracy of the present work and the previous HI$\gamma$S measurement at 61\,MeV provide a strong motivation for further theoretical work on $^4$He in order to extract the neutron polarizabilities with a precision that is difficult to achieve from deuterium experiments.

\section{Summary}
\label{sec:summary}
Elastic Compton scattering from $^4$He provides a complementary approach to deuteron experiments that allows for the extraction of the nucleon polarizabilities. To this end, a new high-precision measurement of the cross section of Compton scattering from $^4$He at 81.3\,MeV has been performed at HI$\gamma$S. While the results exhibit a behavior similar to that seen in previously reported data, this experimental work has achieved an unprecedented level of accuracy. This work, together with the HI$\gamma$S measurement at 61\,MeV, is expected to strongly spur the development of a rigorous theoretical treatment to interpret the $^4$He Compton scattering data in order to extract the polarizabilities of the proton and neutron with high accuracy.

\acknowledgments{
This work is funded in part by the US Department of Energy under Contracts No.~DE-FG02-03ER41231, DE-FG02-97ER41033, DE-FG02-97ER41041, DE-FG02-97ER41046, DE-FG02-97ER41042, DE-SC0005367, DE-SC0015393, DE-SC0016581, and DE-SC0016656, National Science Foundation Grants No.~NSF-PHY-0619183, NSF-PHY-1309130, and NSF-PHY-1714833, and funds from the Dean of the Columbian College of Arts and Sciences at The George Washington University, and its Vice-President for Research. We acknowledge the financial support of the Natural Sciences and Engineering Research Council of Canada and the support of Eugen-Merzbacher Fellowship. We acknowledge the support of the HI$\gamma$S accelerator scientists, engineers, and operators for assisting with the experimental setup, tuning up the accelerator system for high-energy operation, and for the high-quality production of the $\gamma$-ray beam.
}


%

\end{document}